\documentclass[prb,twocolumn,amsmath,amssymb,aps,superscriptaddress,citeautoscript,longbibliography]{revtex4-2}

\usepackage{feynmp}
\usepackage{amsmath}
\usepackage{graphicx}
\usepackage{multirow}
\usepackage{latexsym}
\usepackage{textcomp}
\usepackage{verbatim}
\usepackage{color}
\usepackage{bm}
\usepackage{subfigure}
\usepackage{everysel}
\usepackage{keyval}
\usepackage{ragged2e}
\usepackage{dsfont}
\usepackage{amssymb}
\usepackage{enumitem}
\usepackage{pstricks}
\usepackage{mathtools}
\usepackage{kotex} 

\usepackage{braket}
\usepackage{slashed} 
\usepackage{hyperref}        
\hypersetup{
     colorlinks=true,
     citecolor = cyan,    
     linkcolor=magenta,
     }

\usepackage{graphicx}
\usepackage{xcolor}
\usepackage{amsmath}
\usepackage{bbold}
\usepackage{amsfonts}
\usepackage{bbm}
\usepackage{comment}
\usepackage{lettrine}
\usepackage{orcidlink}

\newcommand{\osum}{{%
    \setbox0\hbox{\circ}%
    \rlap{\hbox to \wd0{\hss\sum\hss}}\box0
}}

\begin{document}

\title{Orbital Edelstein effect of the electronic itinerant orbital motion at the edges}

\author{Jongjun M. Lee$^{\dagger}$\,\orcidlink{0000-0002-9786-1901}}
\thanks{Electronic Address: michaelj.lee@postech.ac.kr}
\affiliation{Department of Physics, Pohang University of Science and Technology (POSTECH), Pohang 37673, Korea}

\author{Min Ju Park$^{\dagger}$\,\orcidlink{0000-0003-1451-0469}}
\thanks{Electronic Address: mandypp@postech.ac.kr}
\affiliation{Department of Physics, Pohang University of Science and Technology (POSTECH), Pohang 37673, Korea}

\author{Hyun-Woo Lee\,\orcidlink{0000-0002-1648-8093}}
\thanks{Electronic Address: hwl@postech.ac.kr}
\affiliation{Department of Physics, Pohang University of Science and Technology (POSTECH), Pohang 37673, Korea}

\begin{abstract}
In the study of orbital angular momentum (OAM), the focus has been predominantly on the intra-atomic contribution. However, recent research has begun to shift towards exploring the inter-atomic contribution to OAM dynamics. In this paper, we investigate the orbital Edelstein effect (OEE) arising from the inter-atomic OAM at the edges. We explore the OAM texture within edge states and unveil the OAM accumulation at the edges using several lattice models based on the $s$ orbital. By comparing slabs with differently shaped edges, we not only clarify the role of electron wiggling motion in shaping OAM texture but also highlight the absence of bulk-boundary correspondence in the accumulation process. The topological insulator and higher-order topological insulator models further confirm these findings and provide evidence for the relationship between the higher-order topology and the OEE. Our study advances the comprehension of orbital physics and extends its scope to higher-order topological insulators.
\end{abstract}

\date{\today}
\maketitle

\section{Introduction}
Studies on electron orbital angular momentum (OAM) in condensed matter are rapidly expanding its scope and significance, unveiling distinct physics and advantages of the electron orbital degree of freedom~\cite{go2021orbitronics}. Despite its foundational importance in physics, OAM in solid systems has received limited attention due to its quenching in equilibrium and its overshadowing by spin angular momentum~\cite{blundell2001magnetism,kittel2018introduction,zutic2004spintronics}. However, recent theoretical and experimental studies are increasingly identifying systems where orbital dynamics emerge as predominant phenomena. For instance, OAM can be induced in centrosymmetric systems under non-equilibrium conditions, giving rise to orbital Hall effect (OHE)~\cite{bernevig2005orbitronics,tanaka2008intrinsic,go2018intrinsic,jo2018gigantic}. This effect not only addresses the problem of overshadowing by spin Hall effect in normal metals but also exhibits universality that extends to various systems, including magnons and phonons~\cite{neumann2020orbital,fishman2022orbital,fishman2023magnon,go2023intrinsic,zhang2014angular,park2020phonon}. Moreover, experimental findings substantiate the presence of OHE and orbital Edelstein effect (OEE) in various materials~\cite{choi2023observation,sala2023orbital,lyalin2023magneto,el2023observation}.

OAM in condensed matter splits into two contributions: intra- and inter-atomic ones. Although both stem from the same underlying physics in a fundamental sense, they are treated separately in modeling and computation. When depicting orbital dynamics such as OHE and OEE, the intra-atomic contribution has been predominantly utilized until now with well-localized $d$ or $f$ orbitals through atom-centered approximation~\cite{go2020theory}. Defining the inter-atomic contribution of OAM on a periodic lattice is non-trivial~\cite{thonhauser2005orbital,xiao2005berry,vanderbilt_2018}. However, a few recent studies incorporated the inter-atomic contribution originated from the itinerant motion across the lattice generalized by the so-called modern theory formulation of OAM. For example, there are several theoretical works on the OHE~\cite{bhowal2021orbital,cysne2022orbital,pezo2022orbital,pezo2023orbital,busch2023orbital} and on the OEE with bulk magnetization~\cite{yoda2018orbital,Leivamontecinos2023spin}. Notably, one recent work by Busch \textit{et al.} demonstrated the occurrence of the OHE in a model involving solely the $s$ orbital. Since the intra-atomic contribution is absent for $s$ orbital systems, OHE should be attributed entirely to the inter-atomic contribution~\cite{busch2023orbital}. This exemplifies a scenario where the atom-centered approximation fails. The authors explained that this unexpected OHE calculated in bulk arises from the OAM generation at the edges of the slab geometry. 

In this paper, we theoretically examine OEE arising from inter-atomic contributions. Previous studies on OEE addressed mostly intra-atomic contributions~\cite{chen2018giant,johansson2021spin,chirolli2022colossal,el2023observation} or inter-atomic contributions of the bulk magnetization~\cite{yoda2018orbital,Leivamontecinos2023spin}. We study the OEE arising at the edges of various two-dimensional $s$-orbital model systems ranging from conventional materials to Chern insulators and higher-order topological insulators (HOTIs). We adopt the tight-binding description to investigate OEE at the edges of these model systems. Initially, we compare OAM textures of straight- and zigzag-edged slab geometries within a square lattice. We observe OAM accumulation exclusively in the zigzag-edged slab, attributed to the wiggling motion of edge states. In contrast, the straight-edged geometry lacks impetus for localized electrons to exhibit such wiggling motion along the edge. We emphasize that this orbital dynamics strongly depends on the edge shape, and there is no bulk-boundary correspondence for this. We concluded it as a manifestation of OEE. We corroborate our findings in the Chern insulator, inherently featuring edge states. Additionally, we show that OEE can typically appear due to the same mechanism in the HOTI. The relationship between HOTI and the orbital dynamics has been raised previously, although direct evidence has been absent~\cite{costa2023connecting}. Our calculations in HOTI models provide direct evidence for the relationship between the higher-order topology and the orbital dynamics explicitly. Our study unveils the intricate interplay between OAM and OEE linking electron motion along edges, and extends insights to HOTI, enriching orbital physics in condensed matter.

This paper is organized as follows: In Sec.~\ref{sec_result}, we present the results of tight-binding calculations conducted on various lattice models of non-interacting electrons characterized by an $s$ orbital. In particular, we investigate a simple square lattice model in Sec.~\ref{subsec_simple_square}, a square lattice model representing the Chern insulator with non-trivial topology in Sec.~\ref{subsec_chern}, and two distinct lattice models of the HOTI in Sec.~\ref{subsec_HOTI}. We demonstrate the OAM accumulation for the models in Sec.~\ref{subsec_OEE}. Our investigation delves into the orbital texture of the localized edge state in the nanoribbon slab, drawing comparisons between different slab geometries and discussing the existence of the OEE in each case. To conclude, we provide a summary of our findings in Sec.~\ref{Sec_conclusion}.

\section{Results and Discussion} \label{sec_result}
Prior to discussing the results, we briefly present the OAM equation in periodic lattice that we will mainly calculate. The OAM in $z$ direction at momentum $k$ for the Bloch eigenstate $|\alpha k\rangle$ of an $\alpha$-th band is given as follows~\cite{thonhauser2005orbital,xiao2005berry,shi2005quantum}.
\begin{equation}
    L^{z}_{\alpha k} = \frac{e\hbar^{2}}{2g_{L}\mu_{B}} \text{Im}\sum_{\beta\neq\alpha} \frac{\langle \alpha k | \hat{v}_{x}|\beta k\rangle \langle \beta k | \hat{v}_{y}|\alpha k\rangle }{\epsilon_{\beta k}-\epsilon_{\alpha k}} ,
\label{Eq_Lz}
\end{equation}
where $-e<0$ is the electron charge, $g_{L}$ is the orbital g-factor, $\mu_B$ is the Bohr magneton, and $\epsilon_{\alpha k}$ is the energy of the $\alpha$-th band. $\hat{v}_{x,y}$ denotes the velocity operator of electrons~\cite{pezo2023orbital,busch2023orbital}. The OAM is a multi-band phenomenon, as Eq.~(\ref{Eq_Lz}) contains a summation of the contraction with all the other bands. Among full descriptions of OAM in a periodic lattice, the above is a self-rotation contribution without considering global rotations of wavepackets, such as the topological edge current. With the formula, we will examine the OAMs for several two-dimensional tight-binding models, especially in their slab geometries with periodicity in a single direction.

\begin{figure}[t!]
    \centering
    \includegraphics[width=8.1cm]{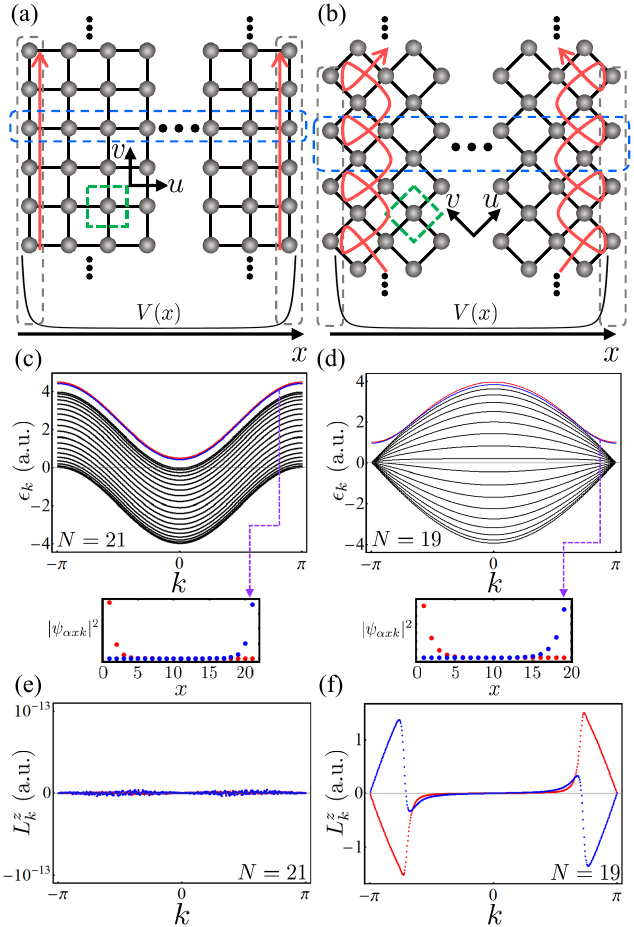}
    \caption{Schematic illustrations of the square lattice model for the slab geometry with (a) the straight and (b) the zigzag edges. The blue and green dashed boxes denote the unit cell of the quasi-one-dimensional slab and the two-dimensional bulk, respectively. The gray dashed boxes and the graph $V(x)$ indicate that the potential is applied at the slab's edges. The red arrows schematically illustrate paths of the electron in the edge state when the electric field is applied from up to down. The energy band $\epsilon_{k}$ for the slab with (c) the straight and (d) the zigzag edges. Localized edge states are colored in red and blue. The insets denote the density $|\psi_{\alpha x k}|^{2}$ along the position $x$ for edge states. The OAM $L^{z}_{k}$, is plotted along the momentum $k$ for the slab with (e) the straight and (f) the zigzag edges, where each color maps to each edge state. The disparity of the shapes between the two plots in (f) arises from a slight imbalance of the on-site energies for the edges. The number of sites in a slab unit cell $N$ is denoted inside each plot.}
    \label{fig1}
\end{figure}

\subsection{Simple square lattice model} \label{subsec_simple_square}
To begin with, we would like to clarify background ideas by investigating a minimal model. We consider a tight-binding model of the simple square lattice [Figs.~\ref{fig1}(a) and (b)], whose nearest-neighbor hoppings are uniform. Then, the Hamiltonian density in momentum space is given by
\begin{equation}
    \mathcal{H}^{\text{sq}}_{\bm{k}}= -2t \sum_{j=u,v} \cos{(\bm{k}\cdot \bm{a}_{j})}c^{\dagger}_{\bm{k}}c_{\bm{k}},
\label{Eq_Ham_sq}
\end{equation}
where $c_{\bm{k}}$ is the electron operator at momentum $\bm{k}$, $t$ is the hopping integral, and $\bm{a}_{j}$ is the lattice vector along $j$ direction. We deal with the lattice in a slab of the ribbon, which is periodic in a single direction, to investigate the edge states. The crystal momentum $k$ is defined along the periodic direction. We compare two geometric conditions: ribbons with straight- and zigzag-shaped edges. They are schematically illustrated in Fig.~\ref{fig1}(a) and (b). We introduce positive on-site energy with a scale comparable to the electron hopping at each side, which can be manipulated by applying gate voltage experimentally. Then, the localized edge states are induced in both geometric conditions, which are well-localized with the momentum near the Brillouin zone boundary. These localized edge states at both sides are distinguished by a slight imbalance in on-site energies between the two sides. The resultant energy bands of the straight and zigzag edges are plotted in Fig.~\ref{fig1}(c) and (d), respectively. Here, the differently colored bands represent the edge states of the opposite sides. The insets of Fig.~\ref{fig1}(e) and (f) denote the wavefunction profile $|\psi_{\alpha x k}|^{2}= |\langle x |\alpha k \rangle|^{2}$ showing the localizations of the edge states in real space.

For both geometric conditions mentioned above, we calculate the OAM of the localized edge states using Eq.~(\ref{Eq_Lz}). The OAM of the edge states from the straight-edge geometry is zero for all points in the momentum space as in Fig.~\ref{fig1}(e), while that from the zigzag-edge geometry is finite as in Fig.~\ref{fig1}(f). It is noteworthy that the two slab geometries, originating from the same bulk Hamiltonian [Eq.~(\ref{Eq_Ham_sq})], demonstrate totally distinct results. This underlines that the existence of well-localized edge states alone does not ensure the occurrence of the OAM texture in the momentum space. In addition to well-localized edge states, a specific edge configuration is also required to manifest the finite OAM texture. For zigzag configuration, electrons go through vortices or rotations at the edge, resulting in finite OAM. Meanwhile, when electrons just go straight, they cannot induce OAM. These paths of the electron are schematically illustrated as red lines in Fig.~\ref{fig1}(a) and (b). The observation that electrons at the opposite edge exhibit opposite signs of OAM in a well-localized regime near the Brillouin zone boundary further supports the above microscopic mechanism of the accumulation. We note that the on-site energy at the edges with a reversed sign also provides localized edge bands below the bulk bands and a consistent result of the OAM texture.

Expanding on the zigzag configuration under an electric field along the $y$-direction, electrons experience a bias in $k$-space, potentially leading to the accumulation of opposite signs of OAM at the edges perpendicular to the applied electric field. This phenomenon is notably absent in the straight configuration. We will explicitly demonstrate this OAM accumulation in the non-equilibrium steady state under the electric field, i.e., the OEE, in this model later in Sec.~\ref{subsec_OEE}. Regarding bulk properties, the orbital Berry curvature is zero within this single-band system, indicating the absence of OHE in the bulk, as determined through the integral of the orbital Berry curvature with the distribution function across the Brillouin zone~\cite{busch2023orbital}. However, discerning OHE, representing the flow of the OAM within a bulk, from OEE poses challenges, as both phenomena can be observed through the accumulation of OAM. The OEE under consideration here is distinct from OHE, as it lacks any correspondence to bulk properties; the accumulation of OAM occurs due to the self-rotation of the localized electrons at the edges. For instance, the phenomenon referred to as OHE in Ref.~\cite{busch2023orbital} may potentially involve a contribution from OEE, which is manifested as orbital accumulation. Moreover, this lack of bulk-boundary correspondence may be related to recent studies on valley~\cite{kazantsev2024nonconservation} and the atomic OAM~\cite{cysne2021orbital,cysne2023orbital} accumulations in nanoribbons.

\subsection{Chern insulator}\label{subsec_chern}
To substantiate the proposed mechanism described in the previous section, we perform similar calculations to the Chern insulator here. In the previous section, we artificially induced edge states in a simple square lattice model by manipulating the on-site potential. In contrast, topological insulators, particularly the quantum spin Hall insulator or the Chern insulator in two dimensions, inherently possess well-localized edge states due to their topological properties \cite{halperin1982quantized,haldane1988model,tong2016lectures}. In this study, we focus on the $\pi$-flux model, a specific type of Chern insulator. While the well-known Haldane model, a tight-binding model with magnetic flux on the honeycomb lattice, could serve as a representative Chern insulator, the $\pi$-flux model was a more suitable choice for our investigation, allowing us to explore and compare edge states in straight- and zigzag-edge geometries \cite{hatsugai2006topological,goldman2013realizing}.

\begin{figure}[t!]
    \centering
    \includegraphics[width=8.1cm]{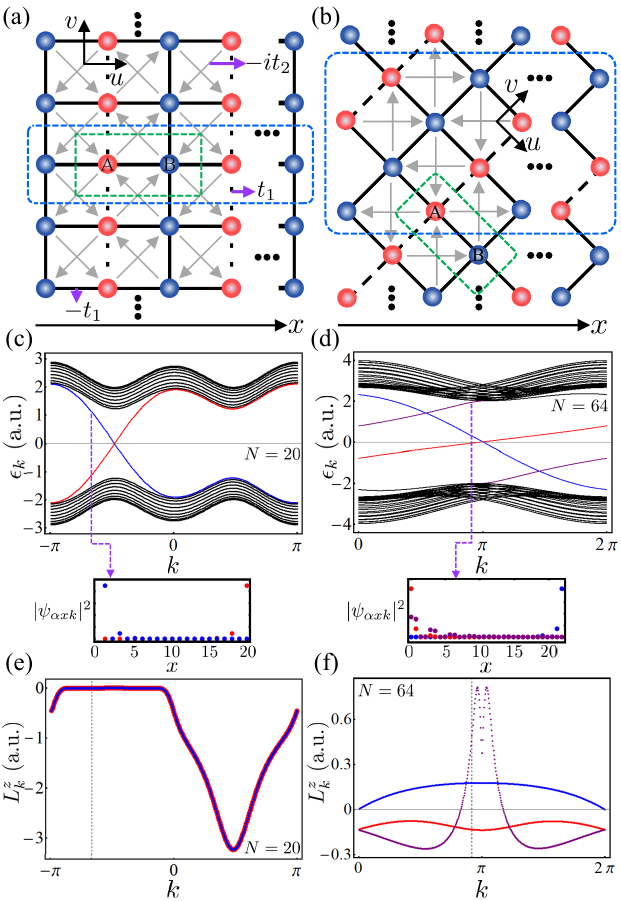}
    \caption{Schematic illustrations of the $\pi$-flux model for the slab geometry with (a) the straight and (b) the zigzag edges. The blue and green dashed boxes denote the unit cell of the quasi-one-dimensional slab and the two-dimensional bulk, respectively. The gray arrow indicates the direction of the next-nearest-neighbor hopping with the phase. The energy band $\epsilon_{k}$ for the slab geometry with (c) the straight and (d) the zigzag edges. Localized edge states are denoted by colors. The insets denote the density $|\psi_{\alpha x k}|^{2}$ at the gray dashed line in (e) and (f) along the position $x$. The OAM $L^{z}_{k}$, which is mapped to each edge state by color, is plotted along the momentum $k$ for the slab geometry with (e) the straight and (f) the zigzag edges. The range of $k$ in (f) is set from $0$ to $2\pi$ for visibility. The number of sites in a slab unit cell $N$ is denoted inside each plot. }
    \label{fig2}
\end{figure}

Hamiltonian density of the $\pi$-flux model in momentum space is given by
\begin{equation}
\begin{aligned}
    \mathcal{H}^{\pi}_{\bm{k}} 
    =& h^{(1)}_{\bm{k}}(a^{\dagger}_{\bm{k}}a_{\bm{k}}-b^{\dagger}_{\bm{k}}b_{\bm{k}}) 
    + h^{(2)}_{\bm{k}}a^{\dagger}_{\bm{k}}b_{\bm{k}}
    + \text{H.c.},
\end{aligned}
\end{equation}
where $h^{(1)}_{\bm{k}}=t_{1}\cos(\bm{k}\cdot \bm{a}_{v})$, $h^{(2)}_{\bm{k}}=-t_{1}(1+e^{-i\bm{k}\cdot \bm{a}_{u}})-2t_{2}\sin(\bm{k}\cdot\bm{a}_{v})(e^{-i\bm{k}\cdot\bm{a}_{u}}-1)$. $a_{\bm{k}}$ and $b_{\bm{k}}$ are electron operators with the momentum $\bm{k}$ at sublattice $A$ and $B$. $t_{1,2}$ is the hopping integral and $\bm{a}_{u,v}$ is the lattice vector along the $u,v$ direction.. The model is visually illustrated in Fig.~\ref{fig2}(a) and (b). Energy bands are calculated for two distinct slab geometries with straight edges and zigzag edges in Figs.~\ref{fig2}(c) and (d). Topologically non-trivial chiral edge states are observed within the gap in both band plots, characteristic of a Chern insulator. This trait, akin to the distinctive features observed in the Haldane model~\cite{haldane1988model}, originates from the complex phase of electrons induced by the imaginary next-nearest neighbor hopping integral, represented as gray arrows in Fig.~\ref{fig2}(a) and (b). The edge states of straight-edge geometry in Fig.~\ref{fig2}(c) are well-localized at the edge when $k\in (-\pi,0)$, as observed in the inset of Fig.~\ref{fig2}(e). Similarly, the edge states of the zigzag-edge geometry in Fig.~\ref{fig2}(d) are well-localized at the edge, as observed in the inset of Fig.~\ref{fig2}(f). However, the purple-colored state at the specific point around $k=\pi$ does not exhibit localization as the others.

We evaluate the OAM of Eq.~(\ref{Eq_Lz}) for the edge states. The corresponding results for the straight-edge and the zigzag-edge geometries are presented in the insets of Fig.~\ref{fig2}(e) and Fig.~\ref{fig2}(f), respectively. The well-localized edge states from the straight edge exhibit zero OAM for both bands, while those from the zigzag edge display nonzero OAMs. This outcome aligns with the findings from the preceding section, suggesting that states from the straight edge do not generate the OAM texture, whereas those from the zigzag edge do. Moreover, in the zigzag configuration under the electric field, the biased electron distribution with an OAM texture is expected to induce OAM accumulation at the edges, a demonstration we will provide in Sec.~\ref{subsec_OEE}. Additionally, we observe that Additionally, we observe that edge states from the red and purple bands are situated on the same edge, unlike those from the blue band. Except for the states near $k=\pi$ of the purple band, which show less localization, states on opposite edges exhibit opposite signs of OAM. This observation strengthens the proposition that the edge shape induces rotations or vortices in the motion of edge electrons. This is because when electrons flow along opposite edges in the same direction, they undergo reversed rotations or vortices. Here, we should remember that these analyses of the slab geometry effects are based on the self-rotation contributions of OAM. Considering intrinsic time-reversal symmetry breaking and resulting Berry curvature and bulk properties, each band has the potential to exhibit a finite value of OAM~\cite{thonhauser2005orbital}. In summary, the results from this more realistic model, which generally possesses the localized edge state, further corroborate the influence of edge shape on the OAM texture and OEE, highlighting a lack of bulk-boundary correspondence analogous to the simple square lattice model.

\begin{figure*}[t!]
    \centering
    \includegraphics[width=17.0cm]{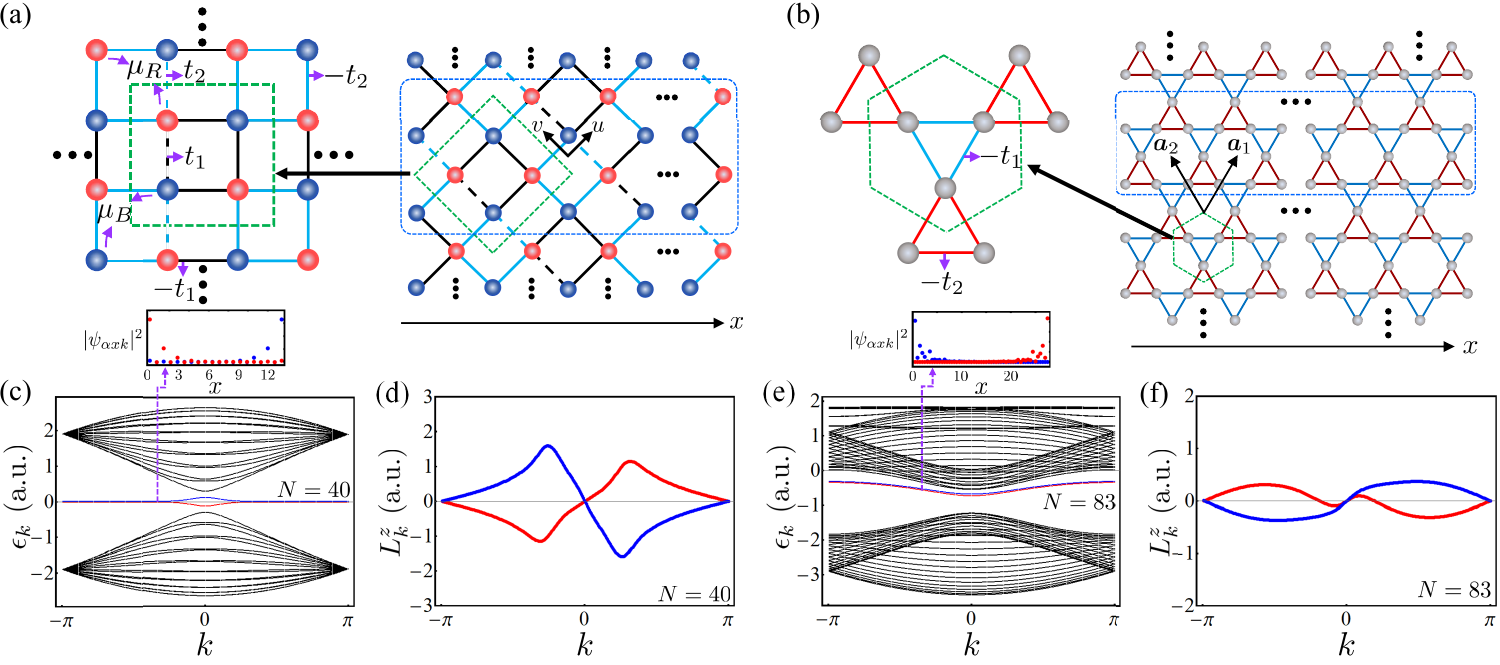}
    \caption{Schematic illustrations of (a) the Benalcazar-Bernevig-Hughes model and (b) the breathing kagome lattice for the tight-binding calculation, respectively. The blue and green boxes denote the unit cell of the slab and the bulk, respectively. The purple arrows denote hopping integrals $\pm t_{1}$ and $\pm t_{2}$ for each hopping path and on-site energies $\mu_{R/B}$ for each sublattice. In (a), the black lines denote intra-cell hoppings, sky blue lines represent inter-cell hoppings, and the dotted lines indicate hoppings with an opposite sign. The energy band $\epsilon_{k}$ of (c) the Benalcazar-Bernevig-Hughes model and (e) the breathing kagome lattice (with the zigzag edges for the Benalcazar-Bernevig-Hughes model), respectively. Localized edge states are indicated in red and blue. The insets denote the density $|\psi_{\alpha x k}|^{2}$ along the position $x$ for each edge state. The OAM $L^{z}_{k}$ of (d) the Benalcazar-Bernevig-Hughes model and (f) the breathing kagome lattice. It is mapped to each edge state by colors and plotted along the momentum $k$. The number of sites in a slab unit cell $N$ is denoted inside each plot. }
    \label{fig3}
\end{figure*}

\subsection{Higher-order topological insulators}\label{subsec_HOTI}
In this section, we investigate the HOTI, with similar approaches employed in previous sections. The orbital physics within HOTI is not yet fully elucidated, though a recent study has reported the HOTI and the OHE in transition metal dichalcogenides~\cite{costa2023connecting}.
HOTI features protected localized states at its hinges or corners, extending the concept of topological insulators. In particular, the second-order two-dimensional HOTI hosts localized zero-dimensional corner states~\cite{benalcazar2017electric,schindler2018higher,xie2021higher}. Various systems have been investigated, revealing numerous models with non-trivial orders. These studies have identified that both the geometric configuration and the electron filling play crucial roles in determining the localization of charge~\cite{benalcazar2017quantized,liu2017novel,ezawa2018higher,ezawa2018higherelectric,park2019higher,song2019all,sheng2019two,xue2019acoustic,mizoguchi2020square,lee2020stable,wakao2020higher,luo2023higher}. For example, the Benalcazar-Bernevig-Hughes model shows quadrupolar localized charges with $e/2$ at each corner of the rectangular sample, and the breathing kagome lattice shows localized charges with $e/3$ at each corner of the triangular sample, where $e$ refers to the charge of an electron~\cite{benalcazar2017quantized,ezawa2018higher}. 
However, in general cases when electron filling or the slab geometry is not appropriate in these systems, the HOTI fails to produce localized corner states, even when the parameter conditions suggest non-trivial topology. In such instances, the localized corner charges usually disperse along the edge, resulting in the emergence of localized edge states. These systems of HOTI with edge states offer a valuable platform for testing our proposed mechanism of the OAM texture, as they inherently exhibit localized edge states in specific geometries and their self-rotating OAM can induce the texture.

We first investigate the Benalcazar-Bernevig-Hughes model, schematically illustrated in Fig.~\ref{fig3}(a). It basically possesses square geometry, so we can compare straight and zigzag edges just as in the previous sections. The Hamiltonian density in the momentum space is given by,
\begin{equation}
    \begin{aligned}
        \mathcal{H}^{\text{bbh}}_{\bm{k}}=& h^{x}_{\bm{k}}(c^{\dagger}_{\bm{k},3}c_{\bm{k},2}+c^{\dagger}_{\bm{k},4}c_{\bm{k},1} )\\
        &+h^{y}_{\bm{k}}(c^{\dagger}_{\bm{k},4}c_{\bm{k},3}-c^{\dagger}_{\bm{k},1}c_{\bm{k},2})+\text{H.c.},
    \end{aligned}
\end{equation}
where $c_{\bm{k},i}$ is the electron operator with the momentum $\bm{k}$ at sublattice $i$. $h^{u,v}_{\bm{k}}=-t_{1}-t_{2}e^{i\bm{k}\cdot\bm{a}_{u,v}}$ where $\pm t_{1,2}$ is the hopping integral, and $\bm{a}_{u,v}$ is the lattice vector along the $u,v$ direction. Additionally, staggered on-site potentials $\pm \mu c^{\dagger}_{\bm{k},i}c_{\bm{k},i}$ is given infinitesimally for calculation simplicity. We calculate the energy band $\epsilon_{k}$ for both straight and zigzag edge slab geometries, each exhibiting well-localized edge states. In Fig.~\ref{fig3}(c), the energy bands of the zigzag slab are presented, featuring colored edge states. The inset of Fig.~\ref{fig3}(d) shows the localization of each state at either side of the edge. Subsequently, we calculate the OAM, $L^{z}_{k}$, for the edge states in both geometries. The edge states of the straight-edge slab consistently exhibit zero OAM across all points in momentum space, akin to previous models, indicating an absence of the OAM texture in this configuration. The outcome is considered trivial and thus not denoted as a figure. In contrast, the edge states of the zigzag-edge slab manifest finite OAM, as plotted in Fig.~\ref{fig3}(d). When an electric field is applied in this case, electrons are expected to exhibit a predominance of either positive or negative signs within $k$-space. Consequently, irrespective of whether the Fermi level resides at the red or blue band, an imbalance in OAM occurs between the opposite edges in this geometry. We will explicitly demonstrate this accumulation later in Sec.~\ref{subsec_OEE}. Here, we note that the bands of edge states become flatter as the width of the slab increases, leading to a decrease in their group velocities. Therefore, for engineering purposes, careful consideration should be given to both the system size and the edge geometry.

We perform a similar analysis for the breathing kagome lattice, which is another well-known example of the HOTI, schematically illustrated in Fig.~\ref{fig3}(b). The Hamiltonian is given by,
\begin{equation}
\begin{aligned}
    \mathcal{H}^{\text{bk}}_{\bm{k}} &= -\sum^{3}_{i=1} (t_{1}+t_{2}e^{i\bm{k}\cdot\bm{b}_{j}})c^{\dagger}_{\bm{k},i+1}c_{\bm{k},i}+\text{H.c.},
\end{aligned}
\end{equation}
where $c_{\bm{k},i}$ is the electron operator with the momentum $\bm{k}$ at sublattice $i$, $c_{\bm{k},i}=c_{\bm{k},i+3}$, $\bm{b}_{1}=\bm{a}_{1}-\bm{a}_{2}$, $\bm{b}_{2}=-\bm{a}_{1}$, $\bm{b}_{3}=\bm{a}_{2}$, $\bm{a}_{1,2}$ is the lattice vector of the Kagome lattice, and $-t_{1(2)}$ is the hopping integral for the intra(inter)-cell hopping. We examine a slab geometry depicted in Fig.~\ref{fig3}(b) with irregular edges. Similarly, we calculate the energy band $\epsilon_{k}$ and the OAM, $L^{z}_{k}$, and the results are plotted in Fig.~\ref{fig3}(e) and (f). The edge states are distinguished by colors, and the inset of Fig.~\ref{fig3}(f) shows that they are well localized at the edges. In this model, the bands of the edge states are almost degenerate with the same signs of the group velocity. The shift of electrons in the momentum space under the electric field results in opposite signs of OAM accumulations on opposite sides of the edge, which is expected to give rise to the OEE.

We emphasize that the OAM texture, and potentially the OEE, broadly arises in the general slab of various HOTI models. Our results provide direct mechanisms and scenarios for the relationship between HOTI and the orbital dynamics, supporting the recent study in transition metal dichalcogenides~\cite{costa2023connecting}.

\subsection{Induced orbital magnetic moment} \label{subsec_OEE}
In the previous three subsections, we have demonstrated that the OAM texture appears in momentum space due to the itinerant orbital motion of electrons at the edges, but only when the edge of the slab geometry is zigzag-shaped. We also briefly discussed the occurrence of the OEE, which involves a non-equilibrium steady state with a tilted distribution in momentum space under an external electric field. Here, we directly demonstrate the OEE, characterized by the accumulation of orbital angular momentum or orbital magnetic moment under an electric field, for the models considered in the previous subsections. This demonstration utilizes the steady-state solution of the semi-classical theory~\cite{cysne2021orbital, cysne2023orbital, Leivamontecinos2023spin, kazantsev2024nonconservation}.

When an electric field $E_{y}$ is applied along the $y$ direction, the distribution function $f_{\alpha k}$ of the $\alpha$-th band at momentum $k$ is modified according to the Boltzmann equation with the relaxation time approximation. The distribution function changes to $f_{\alpha k} = f^{(0)}_{\alpha k} + e\tau ( \partial f^{(0)}_{\alpha k} / \partial \epsilon )|_{\epsilon_{\alpha k}} v_{\alpha k} E_{y}$ up to the linear response. Here, $f^{(0)}_{\alpha k}$ denotes the Fermi-Dirac distribution, $\tau$ is the momentum relaxation time, and $v_{\alpha k}$ represents the group velocity. 

Using this distribution function, one can calculate the spatial distribution of the orbital angular momentum $M_{x}$ by integrating it with the OAM texture, which has been calculated in the previous section, and the weights for the sublattices from the Bloch eigenstate. This is given by the following equation.
\begin{equation}
    M_{x} = M^{(0)}_{ x}
    +\zeta_{x} e\tau E_{y},
\label{Eq_M_x}
\end{equation}
where $M^{(0)}_{x}= \sum_{\alpha, k} f^{(0)}_{\alpha k} |\psi_{\alpha x k}|^{2}L^{z}_{\alpha k}$ and
\begin{equation}
    \zeta_{x} = \sum_{\alpha, k} v_{\alpha k} \frac{\partial f^{(0)}_{\alpha k}}{\partial \epsilon}\Big|_{\epsilon_{\alpha k}} |\psi_{\alpha x k}|^{2}L^{z}_{\alpha k}.
\end{equation}
As indicated by the equations above, the accumulated OAM is proportional to the amplitude of the momentum relaxation time and the applied electric field. The latter term in the equation represents the intrinsic contribution of the induced OAM or orbital magnetic moment, excluding other extrinsic contributions. Its profile should exhibit the OEE, showing their accumulations at the edges.

\begin{figure}[t!]
    \centering
    \includegraphics[width=0.95\linewidth]{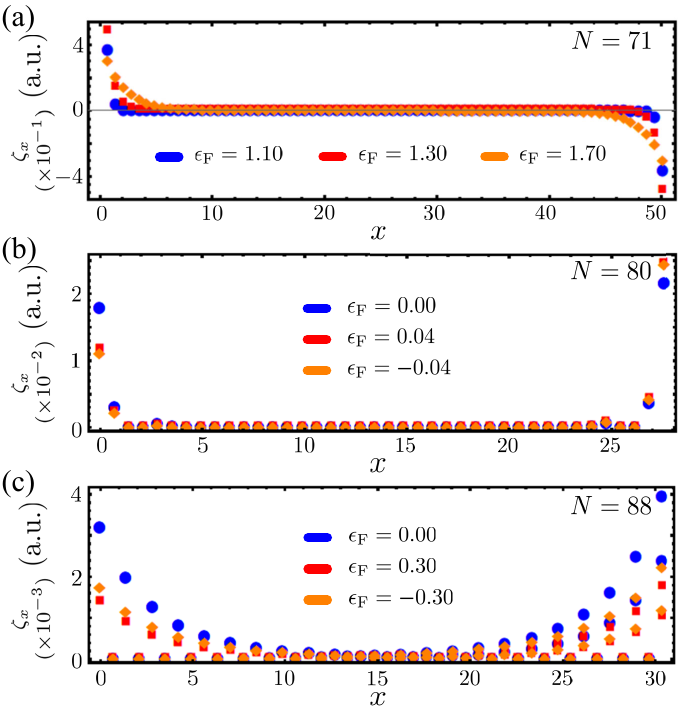}
    \caption{Accumulation of the OAM at the edges induced by the external electric field in (a) the simple square lattice, (b) the $\pi$-flux model, and (c) the Benalcazar-Bernevig-Hughes model. The Fermi energy $\epsilon_{\rm F}$ is controlled. The number of sites in a slab unit cell $N$ is denoted inside each plot.    }
    \label{fig_acc}
\end{figure}

According to the equation, we calculate the OAM accumulation induced by the applied electric field for several models considered in the previous subsections, as shown in Fig.~\ref{fig_acc}. We investigate the contributions of all the edge states and the Fermi energy dependence. All models demonstrate that OAM is accumulated at both edges of the slab geometry. The results of the models can be compared, as the order of the hopping parameter is nearly identical across all models, even though arbitrary units are indicated in the figure. Notably, in the simple square lattice, the accumulation broadens with increasing Fermi energy. This occurs because the increase in Fermi energy results in edge states that are less localized since the wave functions near the Brillouin zone boundary are significantly localized at their low energy level. The accumulation amplitude exhibits a non-monotonic behavior with respect to the Fermi energy since the OAM is maximized at points distant from the zone boundary.

The $\pi$-flux model and the Benalcazar-Bernevig-Hughes model show an even accumulation in $x$, in contrast to the simple square lattice, which shows an odd accumulation in $x$. The sign and even-odd nature of the accumulation are determined by the signs of the group velocity and the OAM at each edge. For instance, applying an electric field to the $\pi$-flux model in a specific direction causes the occupations of the left and right edges to change in opposite directions due to their opposite signs of group velocities. This change in occupation, combining with their opposite signs of OAM, results in an even OAM accumulation. Additionally, the accumulation profiles differ slightly between the left and right sides due to the broken mirror symmetry in the slab geometry. We note that the OHE could arise in topological insulators alongside the OEE during the measurement, potentially generating an additional contribution that further increases the asymmetry of OAM accumulation. Also, the broken mirror symmetry could enhance the OHE contribution, though this depends on the details of the system. The observed OAM accumulation aligns with our expectations and supports the occurrence of the OEE due to the itinerant orbital motion of electrons at the edges. 

Additionally, we note that orbital relaxation is relevant for OAM accumulation at the edges~\cite{seifert2023time,sohn2024dyakonov,idrobo2024direct,rang2024orbital}. For example, depending on the orbital relaxation length, their relative contributions could vary~\cite{nguyen2016spin}. The OEE and OHE may coexist in many systems, but their dependence on the orbital relaxation length differs. While orbital relaxation in itinerant electron motion has not yet been studied and was not explicitly included in our calculations, we expect that incorporating relaxation dynamics would yield more accurate predictions of OAM accumulation at the edges.

\section{Conclusion}\label{Sec_conclusion}
In this work, we investigated the texture of the OAM in momentum space and its associated OEE across various two-dimensional electronic lattice models by using tight-binding calculations. Our findings revealed that the OAM is accumulated at the edges due to the oscillatory motions of electrons prompted by the shape of the sample edges. This aligns with the presumed picture of OAM's origin~\cite{busch2023orbital}. Additionally, our investigation highlighted that this phenomenon lacks correspondence with the bulk properties, and is thus the emergence of OEE rather than OHE governed by the orbital Berry curvature in the bulk. Furthermore, we argued that OEE commonly arises in the HOTI since its localized edge states present OAM textures for the edge state in well-known HOTI models. This observation offers a perspective on the OEE-HOTI relationship~\cite{costa2023connecting}. Finally, our work on $s$ orbital electrons promises to advance the understanding of orbital motion in various other correlated quantum systems, such as the superfluid~\cite{read2011hall,tsubota2013quantum,tada2015orbital}, quantum networks~\cite{chalker1988percolation,medina2013networks,lee2021nonfermi}, etc., in condensed matter physics.

\acknowledgements
J.M.L. and M.J.P. contributed equally to this work$^\dagger$. We acknowledge support from the Samsung Science and Technology Foundation (BA-1501-51). J.M.L. was supported by the POSCO Science Fellowship of the POSCO TJ Park Foundation. We thank Jeonghun Sohn for meaningful discussions.

\bibliography{BibRef}

\end{document}